\documentclass[useAMS,usenatbib]{mn2e}
\usepackage{graphicx}
\usepackage{natbib}
\usepackage{psfig}

\newcommand{\aj}{AJ}

\newcommand{\aap}{A\&A}

\newcommand{\mnras}{MNRAS}

\newcommand{\dmax}{$D_{max}$}
\newcommand{\dmagmin}{$\Delta M_{min}$}
\newcommand{\dczmax}{$\Delta(cz_{max})$}
\newcommand{\dcz}{$\Delta(cz)$}
\newcommand{\dmag}{$\Delta(mag)$}

\title[FGs with XMM]{Characterizing the nature of Fossil Groups with XMM}
\author[F. La Barbera et al.]{
  F.~La Barbera,$^1$\thanks{E-mail: flabarber@gmail.com; labarber@na.astro.it} M.~Paolillo,$^{2,3}$   E.~De Filippis,$^2$ R.R.~de Carvalho$^4$ 
\\
  $^1$ INAF-OACN, via Moiariello 16, I-80128 Napoli, Italy\\
  $^2$Dip. di Scienze Fisiche, Universit\`a di Napoli "Federico II," Compl. Univ. di Monte S. Angelo V. Cinthia, 9, I-80126, Napoli, Italy\\
  $^3$ Istituto Nazionale di Fisica Nucleare, Sez.di Napoli, Italy\\
  $^4$ Instituto Nacional de Pesquisas Espaciais - Divis\~{a}o de Astrof\'{i}sica(CEA), S\~{a}o Jos\'{e} dos Campos, SP 12227-010, Brazil\\
  }

\begin{document}

\date{Submitted on 2011 December 1}

\pagerange{\pageref{firstpage}--\pageref{lastpage}} \pubyear{2011}

\maketitle

\label{firstpage}

\begin{abstract}

We  present an  X-ray follow-up,  based on  XMM plus  Chandra,  of six
Fossil  Group (FG) candidates  identified in  our previous  work using
SDSS  and RASS  data. Four  candidates (out  of six)  exhibit extended
X-ray emission, confirming them as true FGs. For the other two groups,
the RASS  emission has  its origin as  either an  optically dull/X-ray
bright AGN, or the blending of distinct X-ray sources.  Using SDSS-DR7
data, we confirm, for all  groups, the presence of an r-band magnitude
gap between  the seed elliptical and the  second-rank galaxy. However,
the gap value depends, up to  $\sim 0.5$~mag, on how one estimates the
seed  galaxy total flux,  which is  greatly underestimated  when using
SDSS (relative to  Sersic) magnitudes. This implies that  many FGs may
be  actually  missed when  using  SDSS data,  a  fact  that should  be
carefully  taken  into  account  when comparing  the  observed  number
densities of  FGs to  the expectations from  cosmological simulations.
The  similarity   in  the   properties  of  seed--FG   and  non-fossil
ellipticals, found  in our  previous study, extends  to the  sample of
X-ray confirmed FGs, indicating that  bright ellipticals in FGs do not
represent a distinct  population of galaxies. For one  system, we also
find  that the  velocity distribution  of faint  galaxies  is bimodal,
possibly showing  that the  system formed through  the merging  of two
groups.  This  undermines  the  idea  that all  selected  FGs  form  a
population of true fossils.

\end{abstract}


\section{Introduction}

In the late eighties,  Barnes (1989), using simulations, realized that
a merging  of a compact group  of galaxies forms  an elliptical galaxy
over a short fraction of the Hubble time and the end product resembles
a field elliptical. Later, Ponman \& Betram (1993) studying HCG 62 has
concluded that  compact groups result  from orbital decay  of galaxies
into  an extended  dark matter  halo and  the final  system,  a fossil
group, would be  a massive elliptical immersed in  an extended halo of
hot  gas. The  discovery of  such a  system by  Ponman et  al.  (1994)
rendered trustable the whole picture.

Operationally, FGs are systems exhibiting extended X-ray emission with
L$_{\rm X} >  10^{42} {\rm h_{\rm 50}}^{\rm -2} \rm  \, erg \, s^{-1}$
and the  difference in  magnitudes between the  fossil galaxy  and the
second brightest in the group,  $\Delta {\rm m_{\rm 12}}$, equals 2.0,
in  R-band (see  La  Barbera et  al.  2009 for  a  discussion on  this
threshold limit). Only  galaxies picked within a radius  equal to half
of  the virial  radius are  considered. Several  FG samples  have been
defined in  the redshift range 0$<$  z $<$0.6 (Vikhlinin  et al. 1999;
Romer et  al. 2000; Jones  et al. 2003;  Ulmer et al. 2005:  Santos et
al. 2007;  La Barbera et al.  2009; Voevodkin et al.  2010; Aguerri et
al. 2011; Miller  et al. 2011). Although the data  on these systems is
growing fast, their origin is still debatable.  Dariush et al.  (2007)
used the  Millennium Simulation to  show how FGs assembled  their dark
matter halos. They  find that FGs have already their  mass in place by
z$\sim$1  while   their  non-FG  systems  are   still  accreting  mass
today. So,  in this case non-FGs  form later which  is corroborated by
the  work of  D'Onghia et  al. (2005).   This seems  to be  a possible
alternative  to  the compact  group  merging  scenario (Barnes  1989).
However, La Barbera et al. (2009),  examining a sample of 25 FGs and a
similar  control sample  of  non-FGs at  z$<$0.1,  find no  difference
between  structural  and  stellar  population properties  of  FGs  and
non-FGs, suggesting that  FGs may just represent a  transient phase of
mass  assembly in  the Universe.   In fact,  as shown  by simulations,
galaxy  groups  may  undergo  through  a  ``fossil  phase'',  a  stage
characterized by  a large magnitude  gap, terminated by the  infall of
fresh galaxies from the surroundings~\citep{vonbendabeckmann:08}.

In this  work, we  study 6  FG candidates from  our previous  work (La
Barbera  et  al.2009;  hereafter  LdC09),  using  dedicated  XMM,  and
archival  {\it Chandra}  X-ray  observations. This  allows  us to  (i)
assess the success rate of finding true FGs by combining SDSS and RASS
data (as in LdC09); and  (ii) highlight some important pitfalls in the
approaches commonly used  to search for FGs.  We  revisit the issue of
the similarity between seed ellipticals of FGs and non-fossil galaxies
using all  the available data  for these systems, namely,  the density
excess of neighbouring faint  galaxies, distance from the red sequence
in the  color--magnitude diagram, structural parameters  such as $A_4$
(parametrizing the  deviation of galaxy isophotes  from the elliptical
shape,  see e.g.~\citealt{Bender:87})  and  internal color  gradients,
ages, metallicities, and $\alpha$-enhancement.  All these informations
with the X-ray properties, for the first time presented here, form the
basic data set used in the analysis.

This manuscript is organized as  follows: Section 2, discusses how the
sample of FG  candidates was selected; Section 3  presents the new XMM
observations, data reduction and how the X-ray analysis was conducted;
In Section~4 we discuss the main pitfalls of the optical definition of
an  FG, while  in Section~5  we analyze  the velocity  distribution of
group galaxies for  the X-ray confirmed FGs; Section~6  deals with the
number density of FGs, and  the comparison of the properties of fossil
and non-fossil  galaxies. In Section~7 we summarize  the main findings
of this  contribution.  Throughout the  paper, we adopt  the cosmology
$H_0=75    \,    \rm     km  \,  s^{-1} \,   Mpc^{-1}$,    $\Omega_M=0.3$,
$\Omega_{\Lambda}=0.7$.

\section{The Sample}
\label{sec:sample}

The sample  of six  FG candidates studied  here was taken  from LdC09,
where we have  defined a sample of 29  nearby ($z<0.1$) FG candidates,
by combining SDSS  and RASS data.  The selection  of FG candidates was
based  on the  volume-complete  catalogue of  all  galaxies with  r-band
magnitude $M_r  < -20$ and  spectroscopic redshift between  $0.05$ and
$0.095$  from SDSS-DR4. We  defined as  FGs those  elliptical galaxies
with  no companion  brighter  than a  given  magnitude gap,  \dmagmin,
within  a  maximum  projected  radius,  \dmax, and  within  a  maximum
redshift difference,  \dczmax. As a  compromise between the  number of
selected  FG  candidates and  that  of  false  FG detections,  we  set
\dczmax$=0.001$, \dmax $=0.35  \, Mpc$, and \dmagmin $=  1.75 \, mag$.
We notice that these criteria are not the same as those of \dmax $=0.5
R_{vir}$, and \dmagmin $= 2  \, mag$ proposed, on a qualitative basis,
by~\citealt{Jones:03}  (hereafter  J03),  and  widely adopted  in  the
literature   (but  see,  for   instance,~\citealt{Voevodkin:10}).   We
revisit this issue  in Sec.~\ref{sec:fossilness}. We excluded galaxies
with AGN  signatures in the  optical spectra, and those  close (within
$1.5 \, Mpc$) to a rich cluster. Finally, we chose only the 29 systems
with  significant X-ray  emission from  the  RASS, 25  of which  being
classified as  extended relative to  the RASS PSF. All  systems turned
out to have  X-ray luminosity larger than $10^{42}  \, h_{50}^{-2} erg
\, s^{-1}$, i.e.,  the J03 threshold to qualify a system  as an FG. We
have started a  program to observe the 29 FG  candidates of LdC09 with
XMM, to better  characterize their X-ray emission. Because  of the low
S/N and  angular resolution  of the RASS,  the X-ray emission  of each
system requires confirmation through dedicated follow-up observations.
In the present work, we analyze data for six candidates, five of which
classified  as X-ray  extended by  LdC09  (i.e. FG1,  FG3, FG4,  FG12,
FG14), and  the remaining one (FG27) uncertain.  Throughout the paper,
FG candidates are  indicated with the same ID number  as in LdC09 (see
their table~2).  We also compare the properties of seed ellipticals in
the  four  (out  of  six)  confirmed  FGs  with  those  of  non-fossil
ellipticals.  To this effect  we select  the 15  ``field'' ellipticals
brighter  than $M_r=-22.85$ (i.e.  the faint  magnitude limit  of seed
ellipticals in  our sample of  confirmed FGs), out  of the list  of 66
``field'' galaxies with significant X-ray emission presented by LdC09.
We  notice  that 7  (out  of  29) FG  candidates  in  LdC09 have  seed
ellipticals  brighter  than $-22.85$,  making  the  present sample  of
confirmed FGs  representative (although  small) of the  brightest seed
population of FGs.

\section{X-ray data}

\subsection{Observations and data reduction}
\label{data_red}

Only   2  out   of  29   FG  candidates   from  LdC09   had  archival,
good-resolution ($<10``$ FWHM PSF)  X-ray observations. In Cycle 8, we
submitted an XMM proposal to get X-ray data for the remaining FGs. The
proposal was  approved with  priority $\rm C$,  allowing us  to gather
data for four  more candidates. In Table \ref{Xsample}  we present the
final  list of  the available  archival and  proprietary observations.
The proprietary data were obtained by  the EPIC camera on board of XMM
between  September  and  December  2009 (PI  M.Paolillo).   Since  the
observations  were scheduled  with  C priority,  the actual  observing
times exceeded  by a factor of  $\sim 2$ the  requested one.  However,
several of  our observations were affected by  strong flares, reducing
the actual useful data.

The  XMM data (including  archival ones)  were reduced  and calibrated
with the  standard SAS  software v10.0.0; to  maximize the S/N  of the
data we  removed periods affected  by very strong  flares.  Sometimes,
where  flares affected  the whole  observation, we  retained  the data
sections were flares were less severe (see Table \ref{Xsample}). While
in such cases  the data quality is far from  being optimal, it allowed
us to investigate the FG nature  of our targets (see below).  For FG4,
the inspection of the standard {\it Chandra} pipeline products clearly
showed that the X-ray emission  is resulting from a point-like source,
and no additional reduction was performed.

\begin{table*}
\centering
\caption{X-ray observations of our candidate Fossil Groups.}\label{Xsample}
\begin{tabular}{lccrcl}
\hline
id 	& RA, Dec 			& Obs.date 	& Total Exp.time & Removed flaring & Obs.id.\\
	&		 	 	 &			&  (ks)	   	  &	fraction	   & (XMM/Chandra)\\
\hline
FG1	& 13:09:19 --01:37:21 &    2005-06-18	 & 16.9	   &	40\%	   & XMM-0201750201 \\
FG3 	& 07:55:44 +41:12:14 &    2009-10-04   & 16.4	   &	35\%	   & XMM-0605391601 \\
FG4	& 12:53:47 +03:26:30 &   2007-03-04	 & 8.1	   &	--	   & Chandra-8247\\
FG12 & 12:46:51 +00:17:49 & 2009-12-21	 & 9.9	   &	10\%	   & XMM-0605390501 \\
FG14 & 07:51:58 +20:44:56 & 2009-09-30	 & 18.9        &	5\%	   & XMM-0605391501 \\
FG27 & 01:20:23 --00:04:44 & 2009-12-26	 & 7.8	   &	5\%	   & XMM-0605391101 \\
\hline
\end{tabular}
\end{table*}

\subsection{X-ray analysis}

\begin{figure*}
\centering
\includegraphics[width=0.3\textwidth, bb=0 80 595 750]{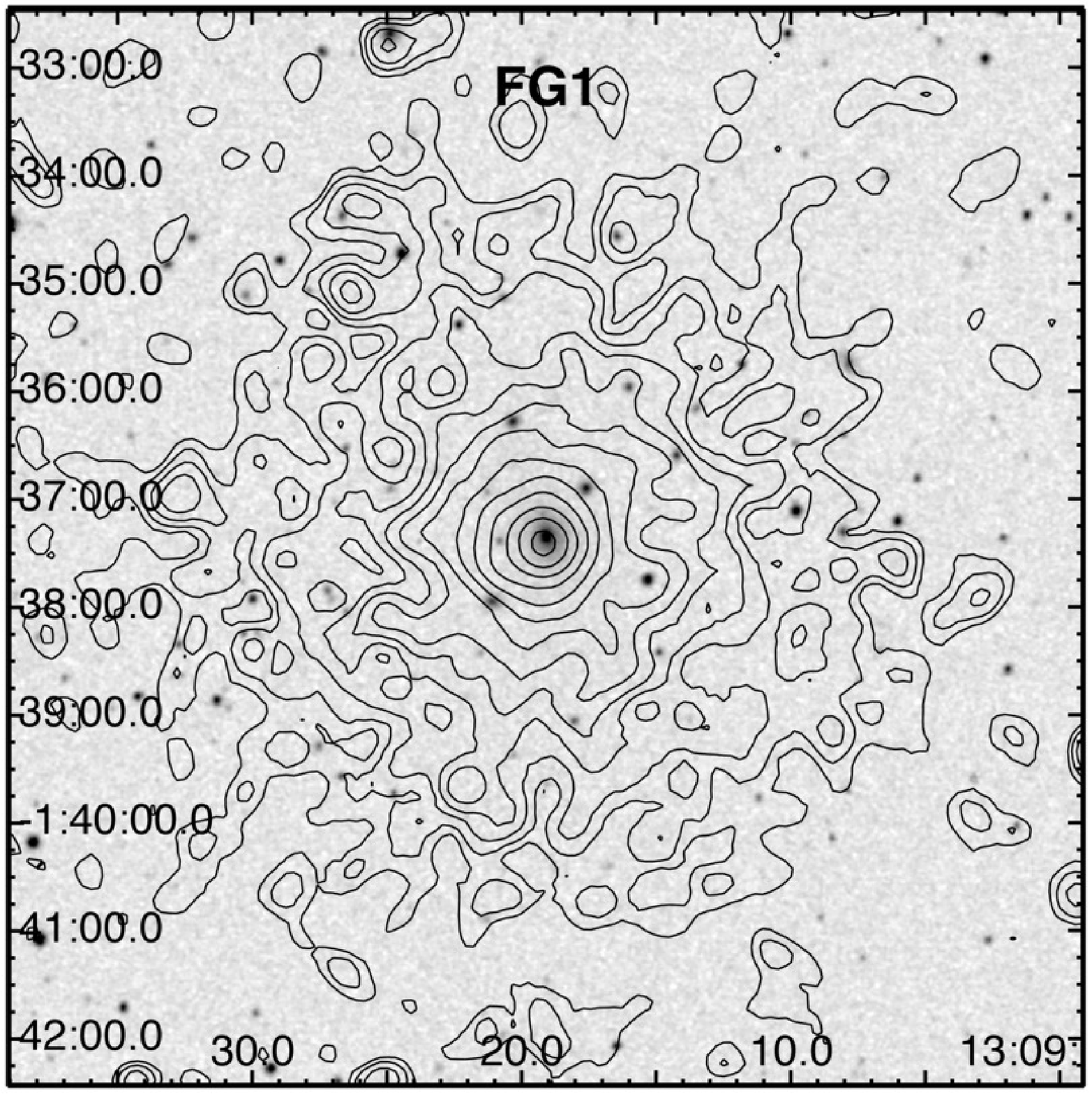}
\includegraphics[width=0.3\textwidth, bb=0 80 595 750]{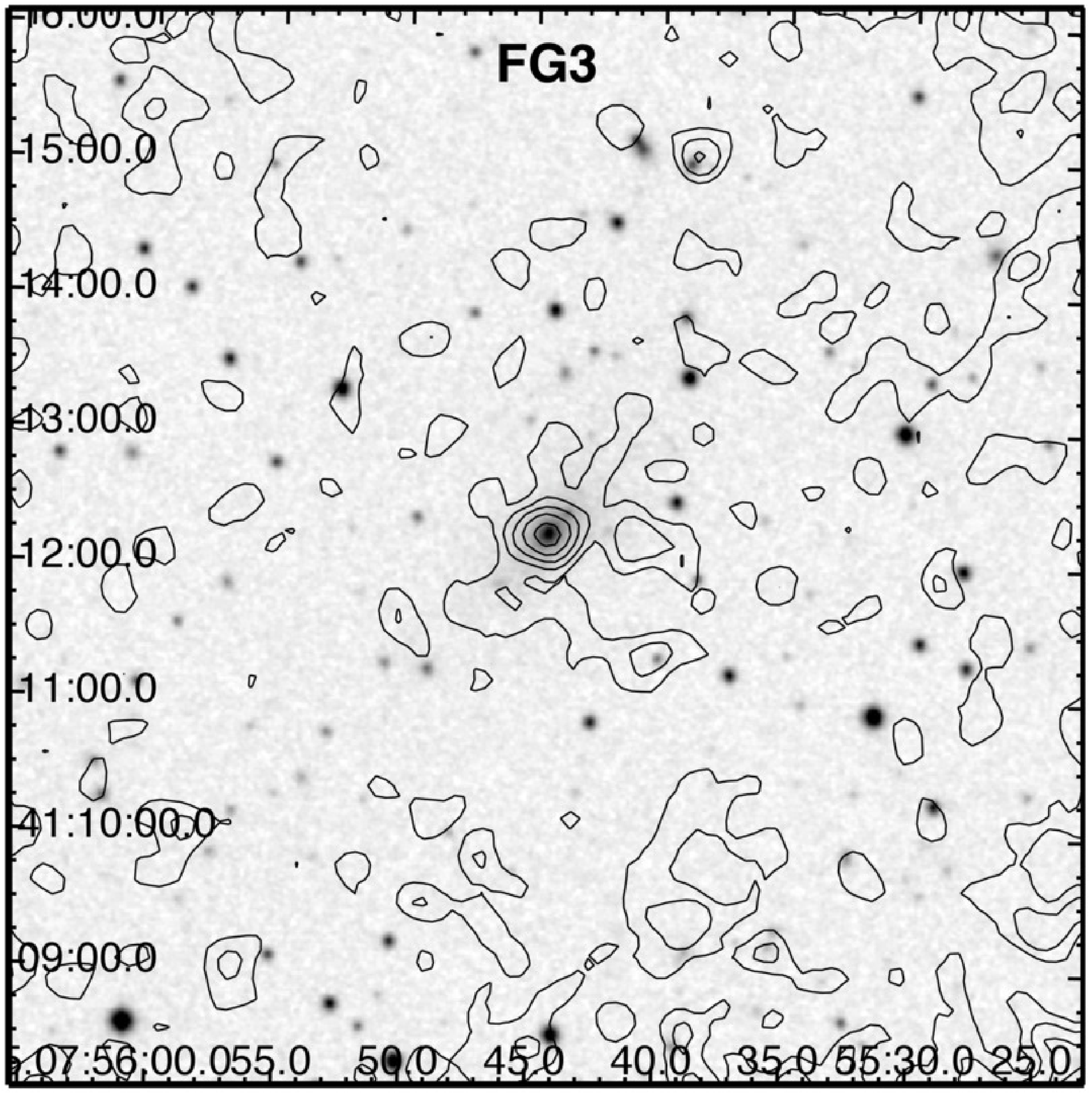}
\includegraphics[width=0.3\textwidth, bb=0 80 595 750]{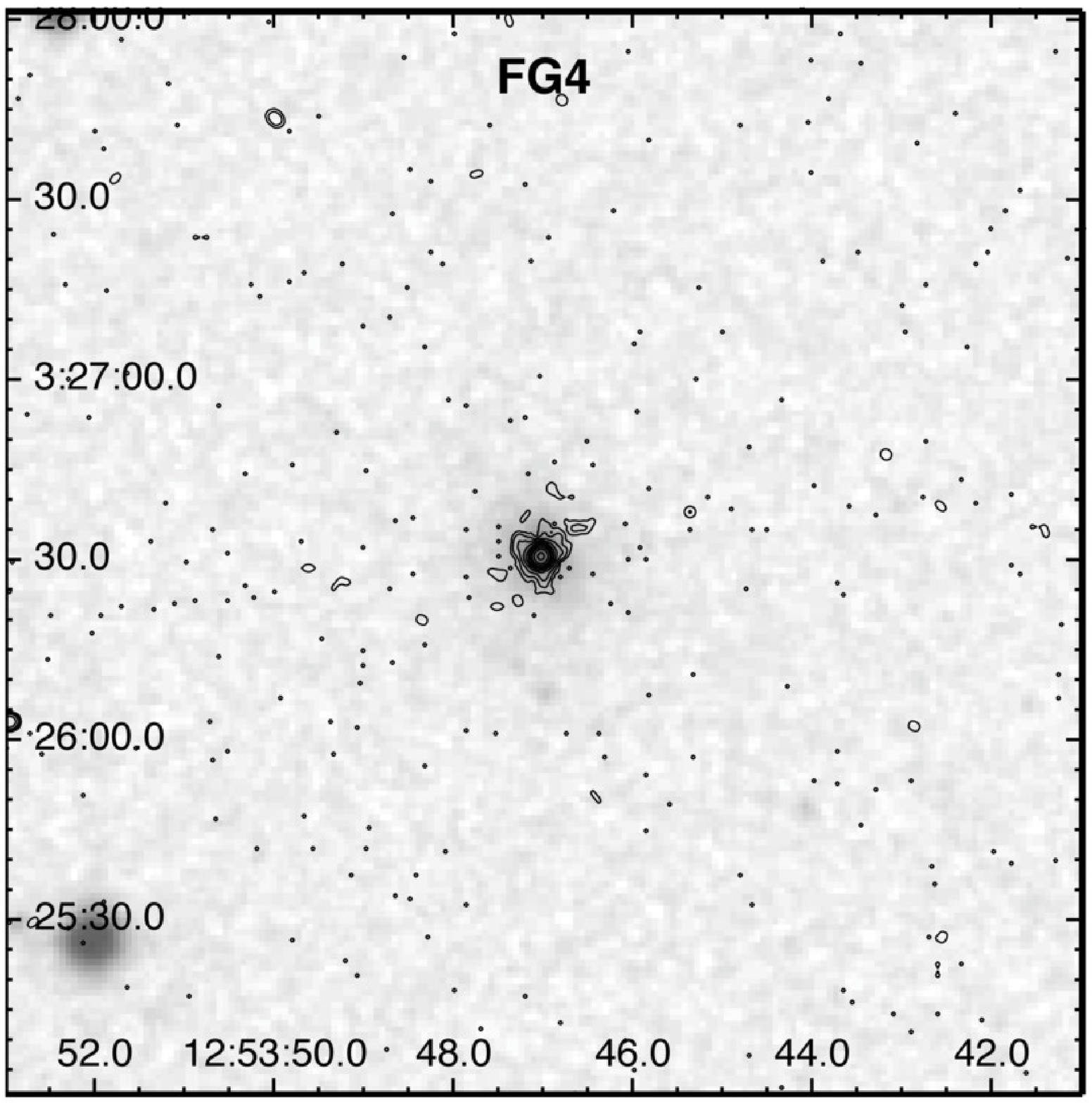}
\includegraphics[width=0.3\textwidth, bb=0 120 595 750]{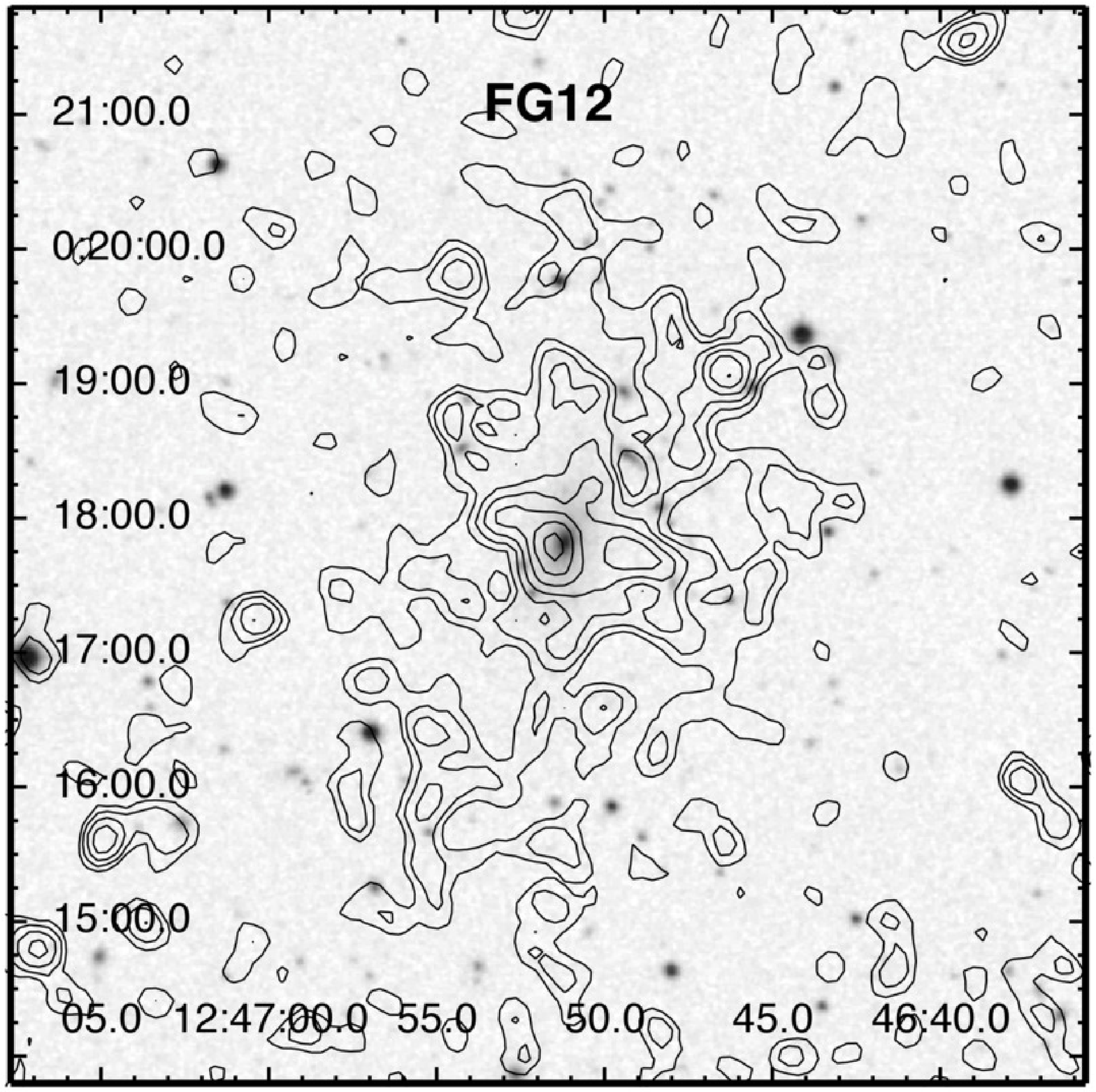}
\includegraphics[width=0.3\textwidth, bb=0 120 595 750]{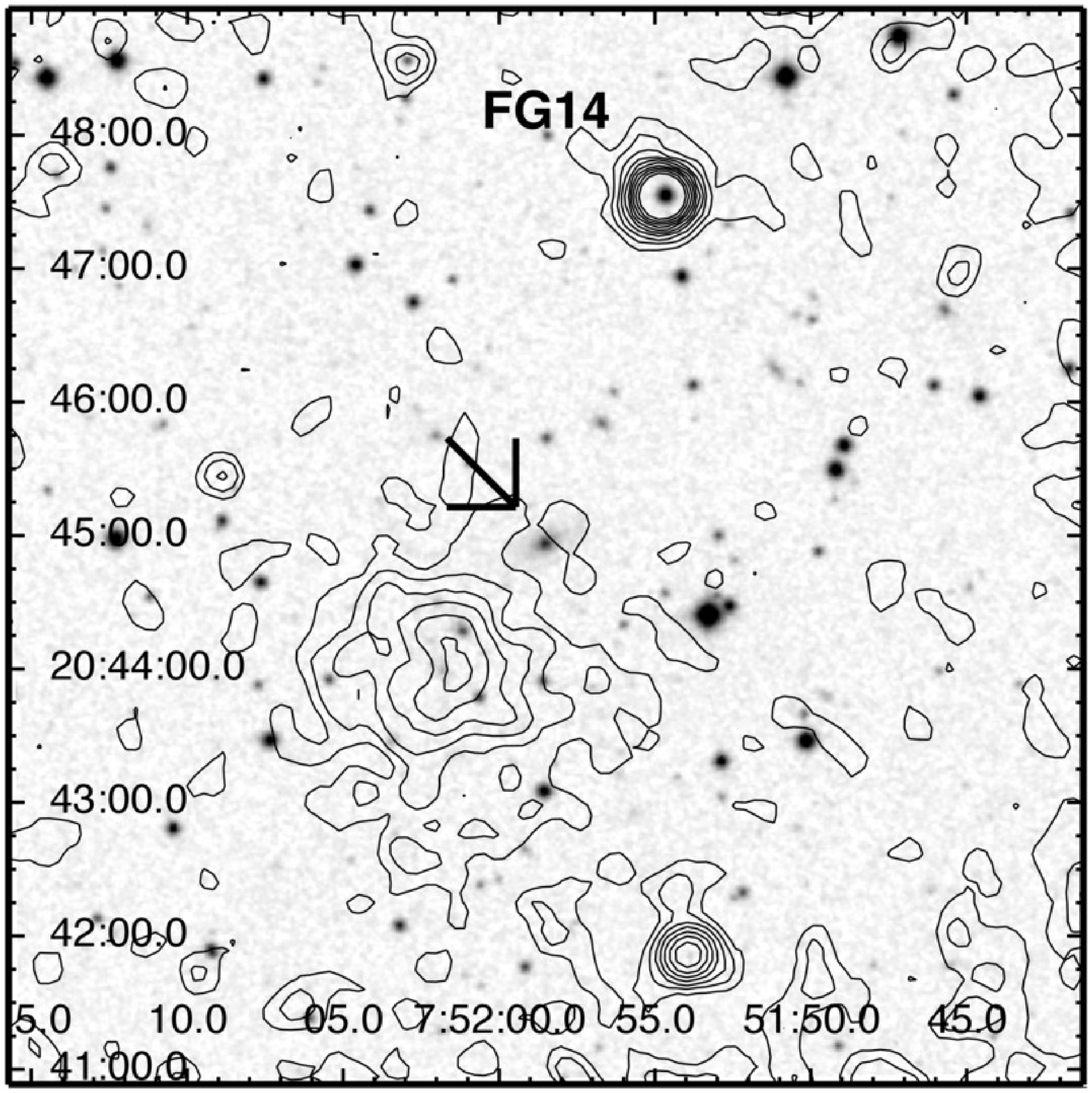}
\includegraphics[width=0.3\textwidth, bb=0 120 595 750]{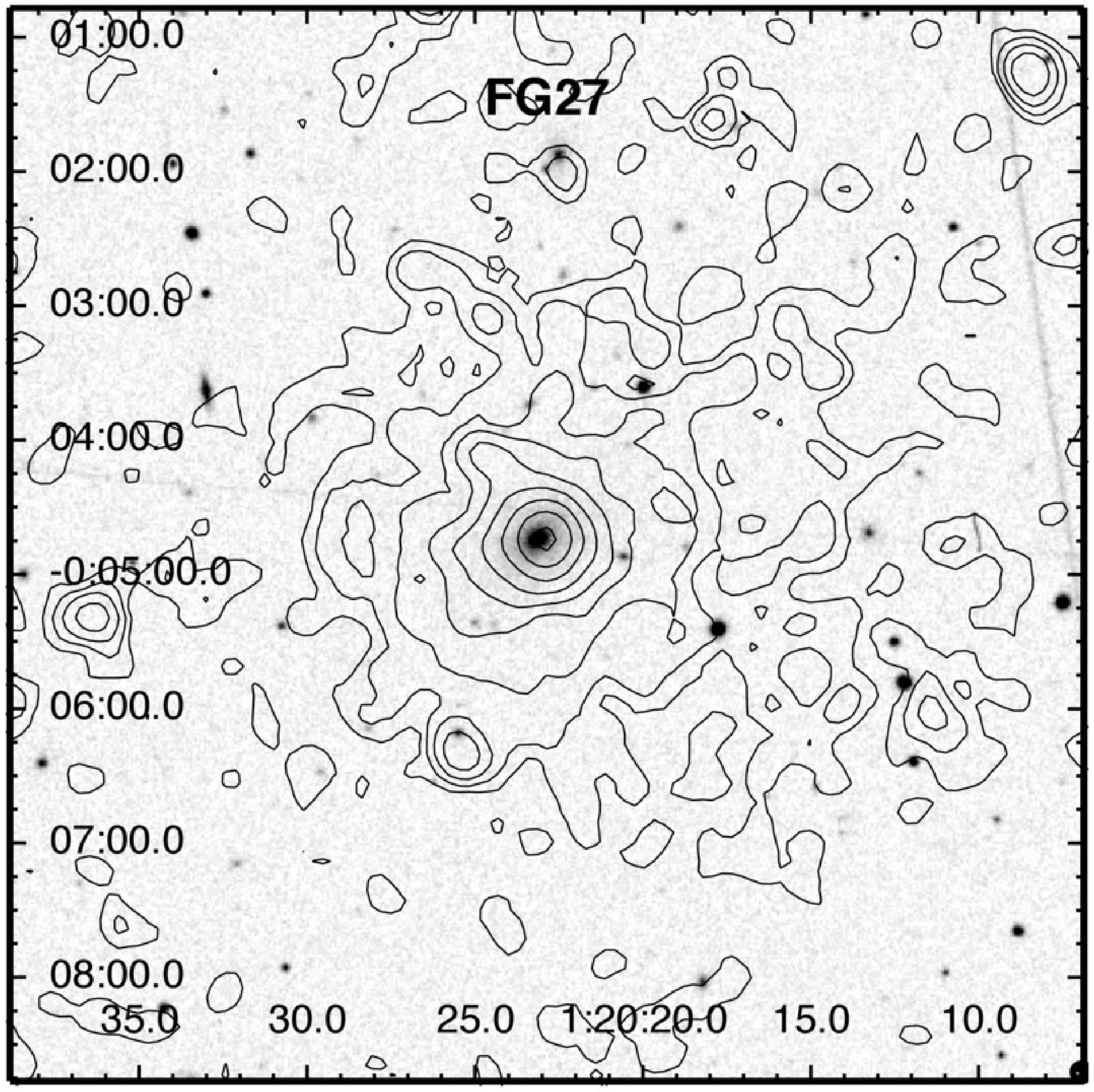}
\caption{X-ray  contours, in the  0.5-2 keV  band, overlaid  on r-band
  DSS2 images. Note that while most images are $8\times 8$ arcmin, the
  field of view  of FG4 is smaller in order  to enhance the visibility
  of the X-ray contours, since for this system the X-ray data are from
  {\it  Chandra} and  then characterized  by a  smaller PSF  and lower
  background. For FG14  an arrow marks the position  of the first-rank
  galaxy,  which does  not correspond  to  the centroid  of the  X-ray
  emission, as discussed in the text.}
\label{X-opt}
\end{figure*}

As a first  step in confirming the FG nature  of our candidate groups,
we extracted  X-ray images  in the 0.5-2  keV band, where  the thermal
emission resulting  from the diffuse  plasma is expected to  peak, for
all sources imaged by XMM. To  this end, we combined the MOS1 and MOS2
detectors.  The  PN  camera  was   not  used  since  it  is  generally
characterized by a higher  background level (especially during flaring
events)  and often  the inter-chip  gap  falls close  to our  targets,
making  an assessment  of the  source extension  more complex.  In all
cases, however, the results of the PN camera agreed with those derived
from the MOS  detectors, although with a lower S/N.  The images of our
targets are shown in Figure \ref{X-opt}.

The images clearly reveal the extended nature of the X-ray emission in
FG1, FG12, and  FG27.  FG3 is fainter and, while  the data suggest the
presence  of extended emission,  a nuclear  component may  account for
part of the X-ray flux~\footnote{However, as discussed in our previous
  work~\citep{Labarbera2009},  this system shows  no AGN  signature in
  the optical spectrum  of its seed galaxy.}.  We  also point out that
the  X-ray  extension   parameter  reported  in  Tab.~\ref{Xprops}  is
actually a lower limit,  considering the relatively high background of
the XMM  data. FG14  represents a unique  case, where there  is indeed
extended X-ray emission, but the X-ray centroid is clearly offset from
the early-type  seed galaxy  of the FG  candidate. In fact,  the X-ray
centroid is centered on  a background cluster, FSVS\_CL J075202+204352
at $z=0.44$  \citep{Sochting2006}, and is  almost certainly associated
with the  latter. The X-ray emission  detected in \cite{Labarbera2009}
is very likely because of  a combination of the background cluster and
two  additional bright AGNs,  visible to  the north  and south  of our
target early-type galaxy.

As  mentioned  in $\S$\ref{data_red},  FG4  only has  \textit{Chandra}
data. In  such case the  inspection of the pipeline  standard products
clearly  shows that  most of  the emission  is produced  by  a nuclear
point-like  source within  the  galaxy  core; i.e.,  an  AGN, with  no
residual extended  emission detectable in the  galaxy outskirts. Since
the selection  of FG  candidates by LdC09  was performed  by excluding
galaxies   with  AGN   signatures  in   the  optical   spectra,  based
on~\citet{BPT} diagnostic diagrams, we conclude that FG4 is one of the
rare cases of optically dull/X-ray bright AGNs, expected to be present
in $\sim 15\%$ of our targets \cite[see e.g.][]{Anderson07,Tozzi06}.

\begin{table*}
\centering
\begin{minipage}{166mm}
\caption{X-ray properties of our candidate Fossil Groups.}\label{Xprops}
\begin{tabular}{lccccccc}
\hline
id 	& redshift	& model & kT 		 	& $n_H$          			&  $Z$ 			& L$_X [0.5-2.0 keV]$            			& extension \\
	&	&            & (keV)	 	& ($10^{20}$ cm$^{-2}$) 	&  ($Z_\odot$)           & ($10^{42}$ erg s$^{-1}$) 	& (kpc)\\
\hline
FG1	&$0.083$	& Mekal & $3.48\pm 0.13$	& $1.2\pm 0.6$	   	& $0.51\pm 0.07$ 	& $18\pm 0.1$ 				& $>460$\\
FG3 	&$0.074$	& Mekal & $1.9\pm 0.6$		& $22\pm 12$ 	   	& [$0.5$]		   	& $2.2\pm 1.1$ 			& $\sim 100$\\
FG4	&$0.066$	& -- 	     & 	--				& 	--   			&	--	   		&  						& pointlike \\
FG12 	&$0.089$& Mekal & $2.5\pm 0.6$		& $1.6\pm 3.4$		& $0.22\pm 0.18$ 	& $3.8\pm 1$ 				& $180$\\
FG14	&$0.077$& -- 	     & 	--				& 	--   			&	--	   		& -- 						& -- \\
FG27	&$0.077$& Mekal & $1.75\pm 0.05$	& $3.8\pm 0.7$	  	& $0.67\pm 0.11$	& $6.8\pm 0.5$ 			& $220$\\
\hline
\end{tabular}
\medskip\\ The extension parameter  refers to the maximum distance out
to which we could detect X-ray  emission. For FG3 we are using EPIC-M1
derived values since  this detector has the best  S/N.  Values between
brackets were  held fixed  in the  fit. FG4 and  FG14 were  not fitted
since there is no detectable hot ISM emission.\\
\end{minipage}
\end{table*}


For the FG candidates  with detectable extended emission, we extracted
spectra from  the 3 XMM detectors  within 200 kpc, and  did a spectral
fit using  the XSpec v12.6  package. FG 4  and FG14 were  not analyzed
since,  as discussed  above, there  is no  evidence of  thermal plasma
filling  the  FG potential  well.   We  used a  Mekal~\citep{MGvdO:85,
  LOG:95} plus photoelectric  absorption spectral model, obtaining the
estimates    of     the    hot    ISM     parameters    reported    in
Table~\ref{Xprops}. The measured  X-ray temperatures and metallicities
are in  agreement with those  expected in galaxy groups  and clusters,
while  the  hydrogen column  density  is  mostly  consistent with  the
galactic value  (see, e.g., Johnson  et al.~2009; Rasmussen  \& Ponman
2007; Mulchaey~2000 and references therein); additional absorption may
be present only  in FG3, but here the uncertainties  are so large (due
to the low  fluxes and large background contamination)  that we do not
consider  this conclusive  evidence  of intrinsic  absorption in  this
system.  While  a  simple  one--temperature  plasma  model  may  yield
sub-solar abundances, we point out that the derived parameters are not
unusual  for  galaxy  groups.  Moreover,  multi-temperature  fits  are
difficult given the low statistics and yield consistent (although very
uncertain) results to those presented here.

In comparing our results with the X-ray luminosity threshold of $L_{X,
  bol}  = 0.44 \times  10^{42} \rm  \, h^{-2}_{75}  \, erg  \, s^{-1}$
adopted by \cite{Jones:03}  to define an FG, we  need to transform the
$0.5$--$2.0$~keV  band into  bolometric  luminosities. The  conversion
factors   lie  in   the  range   $1.8 \rightarrow 2.4$   for   the  different
groups. Hence,  we find that  3 of our  targets (FG1, FG12,  and FG27)
have luminosities  in excess of  the threshold at more  than $3\sigma$
level. For  FG3, the $L_X$ is  still above the threshold,  but only at
the $2\sigma$ level. Finally, we can disprove the FG nature of FG4 and
FG14, since they have no extended X-ray emission.

\section{Fossilness criteria}
\label{sec:fossilness}

In  this Section, we  discuss if  our X-ray  confirmed FGs  follow the
fossilness criteria of J03, where FGs were defined as systems having a
bright  E galaxy,  with no  companion fainter  than $2~mag$  in R-band
within  $0.5 R_{vir}$,  where $R_{vir}$  is the  virial radius  of the
group.  Fig.~\ref{fig:fgpars} shows the  radial and  velocity offsets,
$R$ and \dcz, as a function  of the magnitude difference, \dmag -- all
these quantities being  computed $\sl wrt$ the seed  elliptical -- for
all galaxies in a region  of radius $30'$ around each FG~\footnote{The
  radius of  $30'$ corresponds to $\sim  2.5 \, Mpc$ at  $z \sim 0.08$
  (i.e.   the median  redshift  of the  FG  sample).}. Filled  circles
correspond  to  galaxies with  spectroscopic  redshift available  from
SDSS, within~\footnote{  These galaxies  are likely associated  to the
  FGs, considering that  in each field of view  there are many objects
  within $\pm 2500  \, \rm km \, s^{-1}$ but no  galaxies in the range
  of $2500<|$\dcz$|<4500 \, \rm km \, s^{-1}$.} $\pm 2500 \, \rm km \,
s^{-1}$ to the seed elliptical,  while empty circles are galaxies with
no  spectra from SDSS.  Dotted (dashed)  rectangles correspond  to the
criteria of J03  (LdC09) to define an FG. The  radius of $0.5 R_{vir}$
has been  estimated as in J03,  converting the X-ray  temperature of a
given  FG  (from  Tab.~\ref{Xprops})  into  $R_{vir}$,  by  using  the
relationship   between    X-ray   temperature   and    virial   radius
of~\citet{EMN:96}.  We point out  that the  radial distances  shown in
Fig.~\ref{fig:fgpars} ($R  \le 1.5Mpc$) are larger  that the distances
where we  could detect the  XMM emission, this  being in the  range of
$10\%$ (FG3) to $40\%$ (FG1) of the virial radii.

For all  four systems, no galaxy  is found within  the dashed regions,
i.e. the  systems are certainly fossils  based on the  \dmagmin \, and
\dmax \,  criteria of  LdC09 (see Sec.~\ref{sec:sample}).  Notice that
this is not a trivial result,  as in LdC09, with the aim of minimizing
the  fraction of  FG  detections  by chance,  we  searched for  fossil
systems by considering  only galaxies within $\pm \Delta  cz = 300 \rm
\, km \, s^{-1}$ to the seed  galaxy. Two groups (FG3 and FG12) can be
classified as fossils  also based on the J03 criteria  (as there is no
companion   galaxy  within   the  corresponding   dotted   regions  in
Fig.~\ref{fig:fgpars}).  FG27  would  be  still classified  as  fossil
within $0.5R_{vir}$, if  we slightly relax the  \dmag \, criterium ($\sim
1.9$ instead of  $2$~mags), while for FG1 four  galaxies (two of which
having SDSS  spectroscopy) would invalidate  the J03 definition  of an
FG.  However,  one should  notice that to  date there is  no objective
definition  of what  an  FG is,  and  different criteria  can lead  to
different     samples     of      FGs     (as     discussed,     e.g.,
by~\citealt{Voevodkin:10}). For  instance, the criteria  of LdC09 have
been  introduced  to minimize  the  number  of  spurious (relative  to
genuine) FG detections  from SDSS data, while the  criteria of J03 are
based on  (rather qualitative) physical arguments, e.g.  the fact that
the maximum observed  magnitude gap of galaxy groups  is rarely larger
than \dmag$\sim 1.3  $--$ 1.6$ and $0.5 R_{vir}$  is approximately the
radius within which orbital  decay by dynamical friction predicts that
$M^\star$  galaxies will fall  into the  center of  a group  within an
Hubble  time (assuming  a given  galaxy $M/L$  ratio).   Moreover, one
should notice that classifying a group as fossil may depend on how one
estimates galaxy magnitudes~\citep{Voevodkin:10}, as some galaxies may
enter/exit  the magnitude  gap depending  the way  their  magnitude is
estimated.

To analyze  this issue,  we have selected,  for each FG,  all galaxies
with a (Petrosian)  magnitude~\footnote{We use Petrosian magnitudes as
  they provide the best proxy to  the total flux of a galaxy. However,
  using model magnitudes would not change at all the results presented
  here.} gap ($\sl  wrt$ the seed E) of  \dmag$\le 3$~mags, and within
$1.5$~Mpc to  the position of  the seed galaxies.  We  reprocessed all
these galaxies with the software 2DPHOT~\citep{LdC:08}, fitting galaxy
images  with  2D  seeing-convolved  Sersic  models.   Red  crosses  in
Fig.~\ref{fig:fgpars} show how the  $R$--\dmag \, diagram changes when
using     Sersic,    rather     than    Petrosian     (i.e.     SDSS),
magnitudes. Interestingly,  using Sersic fitting  changes dramatically
the $R$--\dmag  \, plot, making all  four systems to  be classified as
fossils, despite the set of criteria (J03 vs. LdC09) used. This is due
to the  fact that bright ellipticals,  like the FG  seed galaxies, are
described by a high Sersic $n$ profile, for which Petrosian magnitudes
tend   to   underestimate   significantly   the   total   flux   (see,
e,g.,~\citealt{Graham:05}). Since  this effect is  less pronounced for
fainter  galaxies (having  lower $n$,  because of  the luminosity--$n$
relation), using  Petrosian (rather  than Sersic) magnitudes  tends to
decrease the  magnitude gap, hence,  decreasing the number  of systems
classified as fossils.  While this  exercise shows that at least three
of  the four  groups  considered here  (FG3,  FG12, and  FG27) can  be
robustly classified  as fossils  (i.e.  independent of  the fossilness
criteria and magnitude gap definition), it also suggests that many FGs
may be  actually being  missed by currently  available catalogs  of FG
candidates from SDSS (e.g. LdC09, Santos et al.~2007), which are based
on  Petrosian/model magnitudes  to  establish the  magnitude gap  (see
also~\citealt{Harrison:12}).

Recently,   ~\citet{Dariush:10}  (hereafter   DRP10)   suggested  that
\dmag$_{14}$,  i.e. the  magnitude difference  between the  first- and
fourth-rank group galaxy within $0.5  \, R_{200}$, is a more effective
parameter to pick-up early-formed  systems in the Millenium simulation
than \dmag$_{12}$  (i.e.~\dmag \, in our notation  above). A criterium
of  \dmag$_{14} \widetilde{>} 2.5$  would allow  one to  select $50\%$
more groups  with early-assembly epoch  than \dmag$_{12} \widetilde{>}
2$.  From an observation  viewpoint, as  seen in  the lower  panels of
Fig.~\ref{fig:fgpars}, computing  \dmag$_{14}$ is not  trivial, as the
chance of finding  galaxies without redshift from SDSS  is larger when
computing  \dmag$_{14}$  than  \dmag$_{12}$.  For FG27,  we  see  that
\dmag$_{14}$ is $\widetilde{>} 2.5$, independent of the magnitudes one
adopts  (Petrosian  vs. Sersic).  For  FG12,  there  are two  galaxies
without redshift from SDSS  that can significantly affect the estimate
of  \dmag$_{14}$, changing  it  from  $\sim 2.2$  to  $\sim 2.6$  when
considering or not the two galaxies as group members. The same happens
for FG3, while for FG1  one can notice that using Petrosian magnitudes
gives  \dmag$_{14}  \sim  2$,  while  Sersic  magnitudes  would  imply
\dmag$_{14}>2.5$.   In  essence,  one   system  (FG27)   fulfills  the
\dmag$_{14}$ criterium introduced by  DRP10, while for the other three
groups  the  computation  of  \dmag$_{14}$  is hampered  by  the  SDSS
spectroscopic incompleteness as well as the issue of how one estimates
the total flux of the FG seed elliptical.

\section{Velocity distribution of FG galaxies}
\label{sec:velocities}

\begin{figure*}
\centering
\includegraphics[width=17cm]{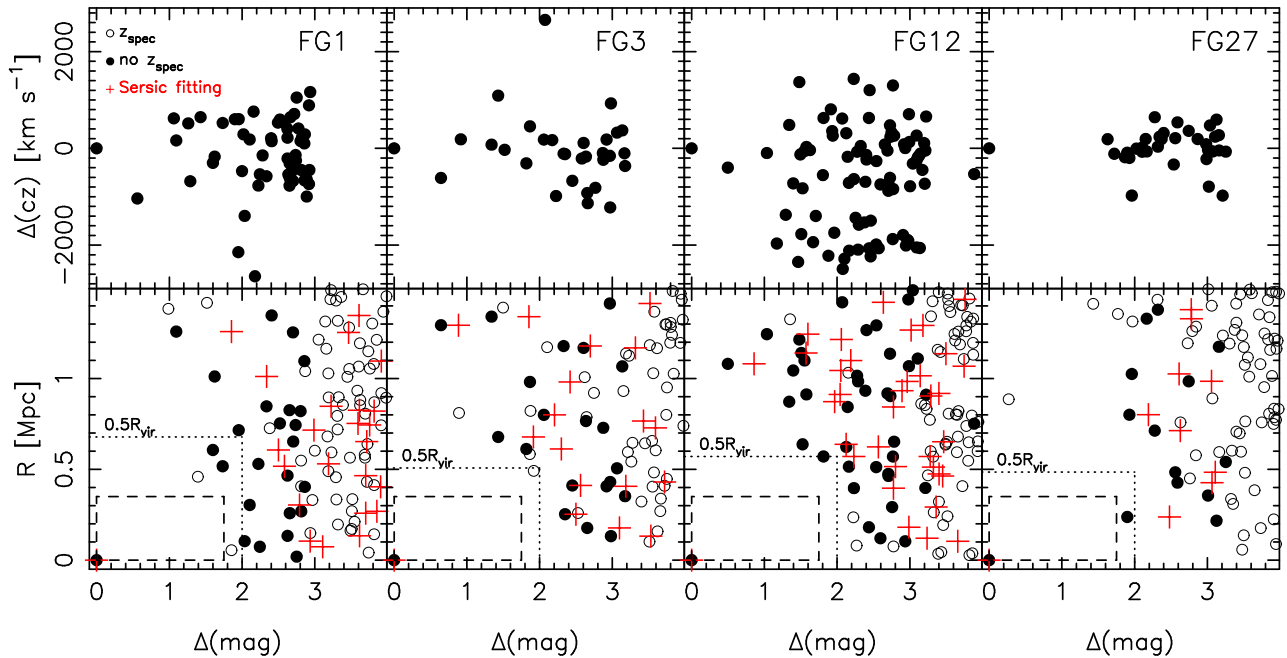}
\caption{Distribution of  the galaxy population of  the four confirmed
  FGs (from  left to right: FG1,  FG3, FG12, and FG27)  wrt their seed
  ellipticals.   Upper  panels plot  the  velocity  offset, \dcz,  and
  r-band magnitude difference, \dmag,  wrt the seed elliptical, of all
  galaxies in the field of view around each FG.  Lower panels show the
  distance to the  seed elliptical (in Mpc) as a  function of \dmag \,
  for all galaxies within $\pm 2500  \, km \, s^{-1}$ from the seed E.
  Filled  (empty)  circles   correspond  to  galaxies  with  (without)
  spectroscopic redshifts  from SDSS-DR7. Red crosses  are obtained by
  selecting all  galaxies with  $R<1.5 \, Mpc$  and \dmag$\le  3$, and
  replacing  their \dmag's  \, with  those estimated  by  fitting each
  object with a seeing convolved Sersic model (see the text). For each
  lower panel,  the dashed  rectangle marks the  region we  adopted in
  LdC09 to search for FGs, while the dotted rectangles are the regions
  defining an FG according to J03.  }
\label{fig:fgpars}
\end{figure*}

\begin{figure}
\centering
\includegraphics[width=8cm]{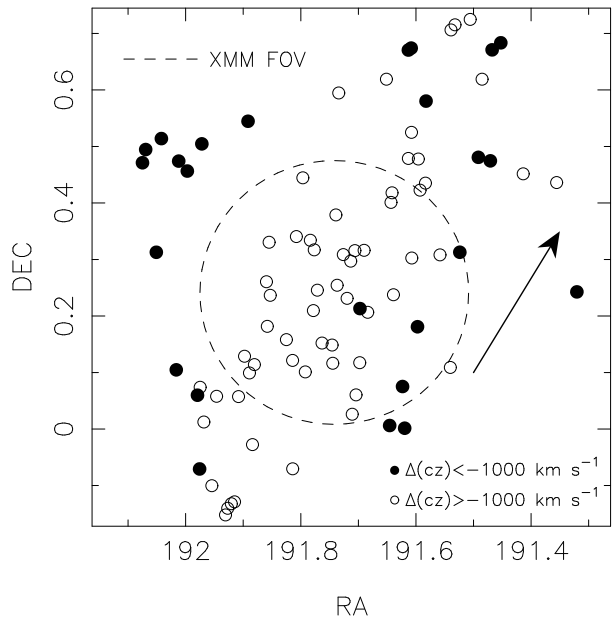}
\caption{Projected spatial  distribution of  galaxies in the  field of
  view of  FG12, within $\pm 2500  km s^{-1}$ and 2.5~Mpc  to the seed
  elliptical.  Galaxies are  split, based on Fig.~\ref{fig:fgpars}, in
  two groups,  those with radial  velocities consistent with  the seed
  galaxy  (\dcz$>-1000  km s^{-1}$;  empty  circles),  and those  with
  \dcz$<-1000  km s^{-1}$  (filled  circles). Notice  that the  latter
  group of  galaxies is  distributed in the  periphery of  the system,
  showing  a  small concentration  at  $RA  \sim  192$ and  $DEC  \sim
  0.5$. Also, galaxies with \dcz$>-1000 km s^{-1}$, i.e. the main body
  of the structure, have elongated shape, in the direction illustrated
  by the arrow.  The dashed circle shows the  field-of-view of the XMM
  EPIC camera.  }
\label{fig:FG12}
\end{figure}

The upper panels of Fig.~\ref{fig:fgpars}  show that for each FG there
are many galaxies with spectroscopic redshift close to the redshift of
the seed  elliptical (within $\pm 2500  \, km \, s^{-1}$  and within a
radius     of     $\sim     2.5$~Mpc).     Using     the     bi-weight
statistics~\citep{Beers:90}, we estimate a velocity dispersion (within
$2.5$~Mpc) of  $\sigma=640 \pm 100$  (FG1), $390 \pm 110$  (FG3), $890
\pm 150$  (FG12), and $280 \pm 50  \, \rm km \,  s^{-1}$ (FG27), where
uncertainties  are  estimated through  $1000$  bootstrap estimates  of
$\sigma$. For FG12, the velocity offset distribution suggests that the
system  has  complex  dynamics,  as  testified by  the  high  $\sigma$
estimate, with some galaxies  having velocity offsets within $\pm 1000
\, km \, s^{-1}$, and the  remaining galaxies being below $-1000 \, km
\,  s^{-1}$.  This  is  quite  surprising, considering  that  FGs  are
expected  to be  evolved systems  that assembled  a large  fraction of
their  mass at  high  redshift (e.g.~\citealt{Donghia:05,  Dariush:07,
  Dariush:10}). Interestingly,  a bimodal velocity  distribution for a
(non-fossil)  galaxy  group ($MZ  \,  4577$)  has  also been  detected
by~\citet{Rasmussen:06}. The  authors associated the  bimodal behaviour
of  the velocity  distribution to  the merging  of two  subgroups, but
found no  convincing evidence that the velocity  substructures had any
spatial segregation on the  sky. Fig.~\ref{fig:FG12} plots the spatial
distribution of  galaxies in the  field-of-view around FG12,  within a
distance of $2.5$~Mpc to the seed galaxy, and absolute velocity offset
$|\Delta(cz)|  < 2500 km  \, s^{-1}$.   Galaxies with  $|\Delta(cz)| <
1000  km \,  s^{-1}$  (filled circles)  define  the main  body of  the
structure, with  a velocity  dispersion of  $500 \pm 90  \rm \,  km \,
s^{-1}$ and  a remarkably elongated  shape, aligned to the  X-ray halo
(see Fig.~1). On  the other hand, galaxies with \dcz$<  -1000 \, km \,
s^{-1}$ (empty  circles) are  preferentially located outside  the main
body of the  structure, with a small aggregation  at $RA \sim 192$~deg
and  $DEC \sim  0.5$~deg, in  the orthogonal  direction $\sl  wrt$ the
elongation  of  the main  structure~\footnote{Selecting  only the  six
  galaxies that define  the small clump at $RA  \sim 192$~deg and $DEC
  \sim 0.5$~deg, we find a velocity dispersion of $200$--$300 \, km \,
  s^{-1}$,  i.e.   significantly  smaller   than  that  of   the  main
  structure.}.   We suggest  that  FG12  might be  the  result of  the
collision  of two  groups of  galaxies that  triggered the  merging of
$L_\star$  galaxies,   leading  to  the  formation  of   the  FG  seed
elliptical. Hence, the faint galaxies we  see at \dcz$< -1000 \, km \,
s^{-1}$ as well as the small  clump at $RA \sim 192$~deg and $DEC \sim
0.5$~deg  would be  just the  relics  of such  collision. The  complex
velocity distribution  of this system  as well as its  elongated shape
reveals that not all the FGs are true fossils, as commonly believed.

In order  to assess how frequent  systems like FG12 are,  we have used
the  group/cluster  sample  defined  by  La  Barbera  et  al.  (2011),
selecting a well  controlled subsample of 82 groups  with similar mass
(0.6 $\times$  10$^{14} \rm M_{\sun}$) and redshift  ($\sim 0.089$) as
for  FG12.  The  mass  for  FG12  was estimated  based  on  the  X-ray
luminosity  listed in  Tab.~2 and  the scaling  relation  presented in
Lopes et al  (2009). We quantified the frequency  of systems like FG12
using the $\sl mclust$ code (using the R language and environment -- R
development  Core Team),  which finds  the number  of  significant and
independent  modes  in the  velocity  distribution.  From the  control
sample  of 82  groups, we  find that  only 12  ($\sim$15\%)  exhibit a
second  component in  the  velocity distribution  as  important as  in
FG12.  The importance  of the  second  component was  measured as  the
relative number of galaxies in the second mode $\sl wrt$ the number of
galaxies in  the main component (i.e.  the one with  largest number of
galaxies) that represents  the group. This can be  done because we are
measuring the number  of galaxies down to the  same limiting magnitude
and within  2.5~Mpc from  the center for  all systems.  If  we further
restrict  to groups whose  two components  have a  redshift difference
$\Delta z \ge 0.006$, which is the value for FG12, then we reduce to 4
groups ($\sim 5\%$). Considering that FG12 is the only of its kind out
of 4 FGs ($25 \%$), and that  in our control sample we find only $\sim
5\%$, or  maximum $\sim$15\%,  it seems that  the dynamical  status of
FG12 is truly linked to its ''fossilness``, namely the merging of two
groups causes the magnitude gap, contaminating a true sample of fossil
systems.   A  larger sample  of  FGs  is  of paramount  importance  to
establish  the   importance  of  such   effect  though~\footnote{In  a
  forthcoming contribution, we plan  to investigate in more detail the
  merging  hypothesis  by  studying   the  properties  of  the  galaxy
  population of FG12}.


\section{Comparison to the findings of LdC09}

\subsection{Number density of FGs}
In  LdC09  we defined  a  sample of  $29$  FG  candidates, $25$  being
classified as X-ray extended according to RASS data, and the remaining
$4$ being uncertain, having an X-ray extension consistent with that of
the RASS PSF. With XMM, we  have observed one of the uncertain systems
($FG27$), concluding  that, indeed, this  is a true FG.  Therefore, we
recompute here the  number density estimate of FGs  using all the $29$
candidates  selected in  our previous  work.   Based on  the XMM  plus
Chandra  data, we estimate  the success  rate of  finding true  FGs as
$f_s=4$(confirmed)/$6$(observed)$\sim67\%$.  At  $L_X  >  0.89  \times
10^{43}  \rm \, h_{75}^{-2}  \, erg  \, s^{-1}$,  where the  sample of
LdC09 is  reasonably (X-ray) complete  (as one can see  by considering
the $L_X$ distribution  of FG candidates from LdC09;  see panel (b) of
their fig.~11), we previously found  21 (out of 29) FG candidates over
the sky area of SDSS-DR4  ($\sim 4873$ square degrees) in the redshift
range $0.05  \le z \le 0.095$,  implying an FG number  density of $7.4
\times 10^{-7} \rm \, h_{75}^3 \, Mpc^{-3}$. Multiplying this estimate
by $f_s$,  we obtain a corrected  number density~\footnote{This figure
  assumes that $f_s$  is independent of $L_X$.  Whilst  this might not
  be  the case,  our number  of confirmed  FGs is  still too  small to
  assess this issue.}  of $\sim 4.9 \times 10^{-7}  \rm \, h_{75}^3 \,
Mpc^{-3}$. Given  an uncertainty of $\sim  30 \%$ in  the success rate
(assuming  binomial statistics  for  $f_s$, see~\citealt{Cameron:11}),
this density estimate  turns out to be significantly  smaller wrt that
found by~\citet{Vikh99}  ($\sim 8.1 \times 10^{-7} \rm  \, h_{75}^3 \,
Mpc^{-3}$) but fully consistent with the expectation from cosmological
simulations~\footnote{Based  on   simulations,  several  studies  have
  estimated  the  expected  number  density  of  fossil-like  systems,
  finding    large    discrepancies    (see~\citealt{Milosavljevic:06,
    Sales:07,  vandenBosch:07}).  In  contrast  to~\citet{Dariush:07},
  these studies analyze optically-defined fossil systems, preventing a
  direct   comparison   to    our   number   density   estimate.}   of
~\citet{Dariush:07}   ($5.4  \times   10^{-7}  \rm   \,   h_{75}^3  \,
Mpc^{-3}$).  We  remark, however,  that  the  sample  of FGs  we  have
confirmed so far  is still small. Moreover, current  FG samples may be
missing  a significant  number  of  systems, because  of  the way  the
magnitude gap is characterized (see Sec.~\ref{sec:fossilness}).

\subsection{Properties of fossil vs. non-fossil ellipticals}

In LdC09, we  have compared several properties of  FG seed ellipticals
with those  of non-fossil galaxies, finding  no significant difference
between  these two  populations.  This  result  applies to  the 25  FG
candidates classified as  X-ray extended based on RASS  data. As shown
in the  present paper,  only some of  these systems are  true fossils,
which could invalidate our previous results. To address this issue, we
compare here the average properties  of brightest galaxies in the four
confirmed FGs  with those of  15 non-fossil ellipticals  selected from
the sample of LdC09 (see Sec.~\ref{sec:sample}).  We consider the main
galaxy properties analyzed by LdC09, i.e. $g-r$ color indices, density
excess  of  faint   companions  around  each  seed/non-fossil  galaxy,
effective  radii and  Sersic  indices~\footnote{Reprocessing the  SDSS
  images of FG galaxies (see Sec.~\ref{sec:fossilness}), we found that
  we previously  underestimated the  Sersic $n$ (effective  radius) of
  the FG1 seed E by $\sim  40 \%$ ($\sim 0.6$~dex).  For a more direct
  comparison with  our previous study, we  use here the  values of $n$
  and $R_e$ presented in tab.~2 of LdC09, but we verified that none of
  the present conclusions would  be altered by updating the structural
  parameters.} (in g- and  r-band), isophotal parameter shape ($A_4$),
internal  optical color  gradients ($\nabla_{g-r}$),  offset  from the
color--magnitude  relation, central  velocity dispersion,  as  well as
stellar population parameters (age, metallicity, $\alpha$/Fe abundance
ratio)  computed  within  the  SDSS fiber  aperture.   For  non-fossil
galaxies,       these      quantities       are       reported      in
Tab.~\ref{tab:OPTprops_noFG}, while  all properties of  FG ellipticals
are reported in tab.~2  of LdC09. Tab.~\ref{tab:MEDprops} compares the
median  values  of all  relevant  galaxy  properties  between the  two
samples.   On average,  we  do not  find  any significant  difference,
although, admittedly, our sample of  confirmed FGs is still small.  We
also list  the p-value for the  two sample t-test  using a permutation
test  ($twotPermutation$  in R  Language),  which  indicates when  two
quantities can  be drawn from the  same parent population.  As one can
see  from   the  Table,  only  $\delta_{CM}$,  the   offset  from  the
color-magnitude relation, shows a tendency to be different between FGs
and  non-fossil ellipticals.   Interestingly, although  the  sample of
non-fossil galaxies  is four  times larger than  that of  fossils, the
error on  the median value  of some galaxy properties  (in particular,
$A_4$  and $\nabla_{g-r}$)  is  larger for  non-fossils, reflecting  a
larger  scatter ($\sl  wrt$  fossils) in  the  distributions of  these
properties. The same  result was also reported by  LdC09 and indicates
that  the formation  of non-fossil  ellipticals is  more heterogeneous
than that  of fossils,  i.e. might involve  a wider range  of physical
processes  (e.g. both dry  and gas-rich  mergers).  This  result seems
consistent with the lack  of star-forming (satellite) galaxies in FGs,
as  recently  found by~\citet{Pierini:11}.   Still  in agreement  with
LdC09, we  find that fossil E's  have on average  boxy isophotal shape
(i.e.  $A_4  <0$), with  FG1 being the  only case with  $A_4>0$ (disky
isophotes).   As already noticed  by LdC09,  this result  contrasts to
that of~\citet{Khosroshahi:06} who found first-rank galaxies of FGs to
have preferentially  disky isophotes, pointing to  gas-rich mergers as
the  driving  mechanism  of  their  formation.   On  the  other  hand,
~\citet{MendezAbreu:11} have recently argued  that dry mergers are the
main drivers of the formation  and subsequent evolution of seed E's in
FGs (see also \citealt{Mendes:09}),  as also suggested by cosmological
simulations~\citep{DiazGimenez08}, more consistent with our findings.

The  result that  fossil and  (some) non-fossil  ellipticals  have, on
average,  similar  (structural   and  stellar  population)  properties
supports a scenario whereby fossils  are just a transient phase in the
life of  galaxy groups~\citep{vonbendabeckmann:08}, and can  be due to
the fact that the magnitude gap  allows one to pick up only a minority
of the early-formed systems~(DRP10).


\begin{table*}
\begin{minipage}{210mm}
\caption{Optical properties of non-fossil ETGs.}\label{tab:OPTprops_noFG}
\tiny
\begin{tabular}{cccccccccccccccccc}
\hline
$RA$ & $DEC$ &$M_r$ & $\Delta M$& $g-r$ &$\delta_N$& $r_{e,g}$&$r_{e,r}$&$n_g$&$n_r$&$A_4$ & $\nabla_{(g-r)}$&$\delta_{CM}$&$\log \sigma_0$&$\log Age$&$[Z/H]$&$[\alpha/Fe]$\\ 
\hline   
$154.93444$ &$ -0.6384$ &$ -23.10 $ & $   2.2 $ & $   0.961 $ & $   3.6 $ & $  1.36 $ & $  1.41 $ & $  4.9 $ & $  4.3 $ & $-0.56 $ & $-0.160 $ & $-1.905 $ & $ 2.455 $ & $ 0.85 $ & $ 0.359 $ & $ 0.360$\\ 
$227.10735$ &$ -0.2663$ &$ -22.86 $ & $   2.3 $ & $   0.975 $ & $   3.3 $ & $  0.42 $ & $  0.57 $ & $  8.4 $ & $  6.0 $ & $-0.29 $ & $ 0.168 $ & $-0.352 $ & $ 2.417 $ & $ 0.86 $ & $ 0.195 $ & $ 0.010$\\ 
$ 22.88717$ &$  0.5560$ &$ -22.85 $ & $   2.1 $ & $   1.015 $ & $   4.6 $ & $  1.63 $ & $  1.37 $ & $  3.9 $ & $  3.5 $ & $ 1.38 $ & $-0.692 $ & $ 1.525 $ & $ 2.438 $ & $ 0.69 $ & $ 0.697 $ & $ 0.330$\\ 
$211.00093$ &$ 64.6827$ &$ -22.88 $ & $   2.8 $ & $   0.946 $ & $   0.3 $ & $  1.15 $ & $  2.04 $ & $  3.4 $ & $ 10.0 $ & $-0.05 $ & $ 1.784 $ & $-1.789 $ & $ 2.504 $ & $ 0.77 $ & $ 0.335 $ & $ 0.290$\\ 
$181.02663$ &$  2.4116$ &$ -22.94 $ & $   2.1 $ & $   0.983 $ & $   4.2 $ & $  0.75 $ & $  0.96 $ & $  5.2 $ & $  6.2 $ & $-0.62 $ & $-0.064 $ & $-0.051 $ & $ 2.395 $ & $ 0.70 $ & $ 0.645 $ & $ 0.290$\\ 
$  3.98205$ &$ 16.2494$ &$ -22.95 $ & $   1.9 $ & $   0.989 $ & $   1.8 $ & $  0.61 $ & $  0.74 $ & $  2.6 $ & $  2.7 $ & $-0.16 $ & $-0.164 $ & $ 0.283 $ & $ 2.483 $ & $ 0.68 $ & $ 0.658 $ & $ 0.290$\\ 
$232.31104$ &$ 52.8640$ &$ -22.93 $ & $   2.2 $ & $   0.995 $ & $   3.2 $ & $  1.33 $ & $  1.39 $ & $  9.8 $ & $  8.6 $ & $-0.34 $ & $ 0.017 $ & $ 0.602 $ & $ 2.454 $ & $ 0.82 $ & $ 0.600 $ & $ 0.410$\\ 
$208.27667$ &$  5.1497$ &$ -23.29 $ & $   2.8 $ & $   1.000 $ & $   6.9 $ & $  1.05 $ & $  0.95 $ & $  6.5 $ & $  5.0 $ & $ 0.91 $ & $-0.121 $ & $ 0.664 $ & $ 2.540 $ & $ 0.64 $ & $ 0.425 $ & $ 0.120$\\ 
$162.30041$ &$ 57.8372$ &$ -22.88 $ & $   2.5 $ & $   0.974 $ & $   2.1 $ & $  2.72 $ & $  1.85 $ & $ 10.8 $ & $ 10.8 $ & $ 2.20 $ & $-0.854 $ & $-0.415 $ & $ 2.447 $ & $ 1.00 $ & $-0.050 $ & $-0.040$\\ 
$135.20756$ &$  4.5833$ &$ -22.85 $ & $   2.8 $ & $   0.977 $ & $   1.5 $ & $  3.00 $ & $  1.55 $ & $  6.4 $ & $  2.3 $ & $-1.24 $ & $-0.838 $ & $-0.215 $ & $ 2.528 $ & $ 0.67 $ & $ 0.822 $ & $ 0.350$\\ 
$131.99603$ &$ 31.7856$ &$ -23.29 $ & $   3.0 $ & $   0.976 $ & $   2.3 $ & $  0.90 $ & $  1.00 $ & $  2.2 $ & $  2.9 $ & $-1.18 $ & $-0.318 $ & $-1.078 $ & $ 2.528 $ & $ 0.89 $ & $ 0.320 $ & $ 0.340$\\ 
$212.51746$ &$ 41.7558$ &$ -23.12 $ & $   3.0 $ & $   0.969 $ & $   3.6 $ & $  1.32 $ & $  1.51 $ & $  6.6 $ & $  6.9 $ & $-0.54 $ & $-0.056 $ & $-1.315 $ & $ 2.477 $ & $ 0.75 $ & $ 0.580 $ & $ 0.220$\\ 
$167.93174$ &$ 40.8207$ &$ -22.90 $ & $   2.6 $ & $   0.901 $ & $   5.4 $ & $  1.61 $ & $  1.78 $ & $  5.8 $ & $  7.5 $ & $ 1.25 $ & $-0.115 $ & $-4.138 $ & $ 2.307 $ & $ 0.71 $ & $ 0.406 $ & $ 0.330$\\ 
$175.19733$ &$ 46.5405$ &$ -22.86 $ & $   2.1 $ & $   0.977 $ & $   0.8 $ & $  0.72 $ & $  0.58 $ & $  3.5 $ & $  3.2 $ & $ 3.42 $ & $-0.184 $ & $-0.270 $ & $ 2.507 $ & $ 0.85 $ & $ 0.290 $ & $ 0.350$\\ 
$133.51888$ &$ 29.0535$ &$ -22.94 $ & $   2.5 $ & $   1.002 $ & $   4.6 $ & $  0.91 $ & $  1.12 $ & $  2.9 $ & $  3.6 $ & $ 1.72 $ & $-0.099 $ & $ 0.964 $ & $ 2.463 $ & $ 0.90 $ & $ 0.350 $ & $ 0.360$\\ 
\hline
\end{tabular}
\medskip\\
\end{minipage}
\end{table*}

\begin{table}
\begin{minipage}{80mm}
\centering
\caption{Median properties of fossil and non-fossil ETGs.}\label{tab:MEDprops}
\begin{tabular}{lccc}
\hline
& seed ETGs & non-FG ETGs & p-value \\
\hline
  $M_r$            &$-23.0 \pm 0.1$     & $  -22.93\pm  0.05$  &    $0.398$ \\    
  $g-r$            &$1.005 \pm 0.015$   & $   0.977\pm  0.01$  &   $ 0.159$ \\    
  $\delta_N$       &$3.0   \pm 0.6$     & $   3.3\pm  0.58 $   &    $0.726$ \\     
  $r_{e,g}$         &$1.14  \pm 0.2$    & $  1.15\pm  0.24 $   &     $0.719$  \\     
  $r_{e,r}$         &$1.04  \pm 0.2$    & $  1.37\pm  0.15$    &     $0.569$  \\      
  $n_g$            &$5.1   \pm 0.9$     & $  5.2\pm  0.85$     &    $0.971$ \\       
  $n_r$            &$4.8   \pm 0.8$     & $  5.0\pm  0.88 $    &    $0.547$ \\      
  $A_4$            &$-0.34 \pm 0.2$     & $-0.16\pm  0.44$     &    $0.374$ \\       
  $\nabla_{(g-r)}$   &$-0.1  \pm 0.07$   & $-0.121\pm  0.20$    &    $0.953$     \\      
  $\delta_{CM}$     &$1.1   \pm 0.8$    & $-0.270\pm  0.45 $   &    $0.117 $ \\     
  $\log \sigma_0$  &$2.5   \pm 0.04$    & $ 2.463\pm  0.02 $   &   $ 0.494$  \\     
  $\log Age$       &$0.75  \pm 0.08$    & $ 0.77\pm  0.03$     &   $ 0.564$ \\       
  $[Z/H]$          &$0.38  \pm 0.05$    & $ 0.406\pm  0.07$    &   $ 0.576$  \\      
  $[\alpha/Fe]$    &$0.25  \pm 0.05$    & $ 0.330\pm  0.04$    &   $ 0.814$  \\      
\hline
\end{tabular}
\medskip\\
\end{minipage}
\end{table}




\section{Summary}
\label{sec:Summary}

We  have presented X-ray  data for  six FG  candidates defined  in our
previous  work  as systems  with  significant  magnitude  gap, in  the
optical,  between   first-  and  second-rank  galaxies,   as  well  as
significant X-ray emission from RASS. We  confirm 4 out of 6 groups as
true    FGs,   with   extended    X-ray   emission    and   luminosity
$L_X[0.5$--$2.0keV]  \widetilde{>}  2  \cdot  10^{42} \,  \rm  erg  \,
s^{-1}$, implying  a success  rate of $\sim  67\%$ when  searching for
these  systems   with  SDSS+RASS.  Contamination   mainly  comes  from
optically dull/X-ray  bright AGNs  (expected to be  $\sim 15\%$  ), or
blending of different X-ray sources around the target elliptical.

We analyze the distribution of galaxies around each confirmed FG, and compare the properties of (confirmed) fossil and non-fossil ellipticals selected from our previous work (LdC09). The main results are the following:

\begin{description}
 \item[-]  For   all  systems,  we   find  many  galaxies   with  SDSS
   spectroscopic redshift close to that of the seed elliptical, within
   a maximum  distance of $\sim  2.5$~Mpc. The velocity  dispersion of
   FGs ranges from $\sim 300$ to $\sim 900 \, \rm km \, s^{-1}$.
 \item[-]  All FGs  exhibit a  clear  magnitude gap,  \dmag, within  a
   maximum radius, \dmax, from the seed galaxy. However, the values of
   \dmag \,  and \dmax \, critically  depend on how the  total flux of
   the  seed  galaxy  is  estimated. Being  bright  ellipticals,  seed
   galaxies have a surface brightness distribution well described by a
   Sersic  profile with  high  $n$ ($>4$),  implying that  Petro/model
   magnitudes  from SDSS tend  to underestimate  their total  flux (by
   $\sim 0.5$~mag), and hence the magnitude gap. Considering that most
   works use a  sharp value of \dmag \  (typically $2$~mags) to search
   for  FGs, our  finding means  that current  SDSS-based  samples may
   actually miss  a significant fraction of FGs.  In particular, using
   \dmag$=2$  and SDSS  magnitudes,  we would  lack $1$--$2$  systems,
   i.e. $25$--$50 \%$, out of 4  FGs in our sample. This effect should
   be taken  into account  when comparing FG  number densities  to the
   expectations of cosmological simulations, potentially unaffected by
   such bias.
\item[-]  We restrict  the comparison  of properties  of  seed--FG and
  non-fossil  ellipticals to  the sample  of X-ray  confirmed systems,
  finding no  significant difference  between the two  populations, in
  agreement  with LdC09.  The only  remarkable feature  is  the larger
  scatter  of  non-fossil  (relative  to  fossil)  galaxies  for  some
  properties (i.e. internal color  gradients and $A_4$ isophotal shape
  parameter).  We do not  find FG  ellipticals to  have preferentially
  disky  isophotes,  i.e., they  do  not  form preferentially  through
  gas-rich mergers, as previously claimed in the literature.
\item[-] Considering  our success rate  to find true FGs,  our current
  number  density  estimate for  these  systems  is  $\sim 4.9  \times
  10^{-7} \rm \, h_{75}^3 \, Mpc^{-3}$ (for $L_X > 0.89 \times 10^{43}
  \rm \,  h_{75}^{-2} \, erg  \, s^{-1}$), fully consistent  with what
  expected   from  cosmological   simulations~\citep{Dariush:07},  but
  smaller $\sl wrt$ the independent estimate of~\citet{Vikh99}.
\item[-]  Lastly,  we find  that  galaxies  around  one of  the  X-ray
  confirmed   FGs   (FG12)  shows   a   remarkably  complex   velocity
  distribution. Most galaxies have  velocities within $\pm 1000 \rm \,
  km \, s^{-1}$  to the seed elliptical, defining  a clearly elongated
  spatial   distribution,   approximately   parallel  to   the   X-ray
  emission. A secondary group  of galaxies has average velocity offset
  of about $-2000  \rm \, km \, s^{-1}$ to the  seed. They are located
  more than  $1$~Mpc ($\sim 1  R_{vir}$) far from the  main structure,
  with a clumpy spatial distribution.  We argue that FG12 might be the
  result of the  merging of two groups, that  triggered the merging of
  $L_{\star}$  galaxies,   leading  to  the  formation   of  the  seed
  elliptical. This  suggests that  not all FGs  are truly  fossils, as
  commonly thought.
\end{description}

The present work points to the importance of next generation wide-area
X-ray telescopes,  such as eRosita  (Cappelluti et al. 2011)  and WFXT
(Rosati et al.2011), to obtain large samples of FGs, allowing a robust
comparison with predictions from cosmological simulations.

\section*{Acknowledgments}
We  are thankful  to  A.~Dariush, E.~D'Onghia,  and A.~Finoguenov  for
helpful comments and suggestions.  MP and FLB acknowledge support from
the ASI-INAF  contract I/009/10/0.  MP also acknowledges  support from
MIUR PRIN-2009.   This paper  uses data of  SDSS-DR7. Funding  for the
SDSS and SDSS-II has been  provided by the Alfred P.~Sloan Foundation,
the Participating  Institutions, the National  Science Foundation, the
U.S.   Department  of  Energy,  the  National  Aeronautics  and  Space
Administration, the  Japanese Monbukagakusho, the  Max Planck Society,
and the  Higher Education Funding  Council for England.  The  SDSS Web
Site   is  http://www.sdss.org/.    The   SDSS  is   managed  by   the
Astrophysical Research Consortium  for the Participating Institutions.
The  Participating Institutions  are  the American  Museum of  Natural
History,  Astrophysical   Institute  Potsdam,  University   of  Basel,
University of  Cambridge, Case Western  Reserve University, University
of Chicago,  Drexel University,  Fermilab, the Institute  for Advanced
Study, the  Japan Participation  Group, Johns Hopkins  University, the
Joint  Institute for  Nuclear  Astrophysics, the  Kavli Institute  for
Particle Astrophysics  and Cosmology, the Korean  Scientist Group, the
Chinese Academy of Sciences  (LAMOST), Los Alamos National Laboratory,
the     Max-Planck-Institute     for     Astronomy     (MPIA),     the
Max-Planck-Institute   for  Astrophysics   (MPA),  New   Mexico  State
University,   Ohio  State   University,   University  of   Pittsburgh,
University  of  Portsmouth, Princeton  University,  the United  States
Naval Observatory, and the University of Washington.



\label{lastpage}

\end{document}